\documentstyle[11pt,aaspp]{article}

\begin {document}

\title{The Discovery of Two FU Orionis Objects in L1641}
\author{Karen M. Strom\altaffilmark{1} and Stephen E. Strom\altaffilmark{1}}
\affil{Five College Astronomy Department, Graduate Research Center,517-G,\\
University of Massachusetts, Amherst, MA 01003\\
Electronic Mail: kstrom@hanksville.phast.umass.edu;
sstrom@donald.phast.umass.edu}

\received{1993 April 5}

\altaffiltext{1}{Visiting Astronomer, Kitt Peak National Observatory,
National Optical Astronomy Observatory, which is operated by the Association
of Universities for Research in Astronomy, Inc. (AURA) under cooperative
agreement with the National Science Foundation.}

\begin{abstract}

We have obtained spectra of the reflection nebulosity illuminated by two
heavily embedded IRAS sources in the L1641 molecular cloud, Re 50 and IC
430 (V883 Ori). Examination of these spectra in combination with their
other properties indicates that these objects are FU Orionis objects.
Our spectrum of L1551 IRS 5 confirms the FU
Ori classification made for this star by Mundt et al. (1985).

\end{abstract}

\section{Introduction}

Over the last two decades extensive optical/infrared studies of young
optically visible pre-main sequence stars have helped to establish a
paradigm for the later stages of the star formation process. To date the
properties of their precursors have been inferred largely from
photometric measurements.
We have begun a program to examine the spectra of heavily embedded
pre-main sequence objects by taking advantage of the fortuitous
circumstance that some of these objects illuminate reflection
nebulae. In Lynds 1641 reflection nebulosity is
associated with several of the heavily embedded IRAS sources that have no
other optical manifestation. The two objects which exhibit the highest
surface brightness nebulosity at R are IC 430 (Haro 13a, V883 Ori, 05358-0704),
19.8
mag/\sq\arcsec, and Re 50 (05380-0728), 20.8 mag/\sq\arcsec.

IC 430 appeared in a list of nebulous objects in the Orion region contained in
the H$\alpha$ survey of Haro (1953). While H$\alpha$ emission was not detected
from this object, the morphological similarity to HH objects led Haro
to suspect association with a star formation event. A faint star, below
Haro's plate limit, is seen at the tip of the nebula and has since been
designated V883 Ori. This star was interpreted as an embedded middle B
main sequence star by Allen et al. (1975) on the basis of an almost featureless
image tube
spectrogram in the H$\alpha$ region, which showed neither H$\alpha$ emission
nor absorption, and on its absolute K magnitude.

Re 50 (Reipurth 1985a) was first discussed by Reipurth and Bally (1986) as
the source of a powerful molecular outflow whose associated reflection
nebulosity
has only recently appeared in the blue lobe of the outflow. Polarization
studies (Scarrott \& Wolstencroft 1988) and near infrared imaging (Casali 1991)
have demonstrated that the nebula is illuminated by the IRAS source.

\section{Observations}

On the night of 1993 January 26 UT we obtained spectra of Re 50 and IC 430
using the RC Spectrograph on the 4 meter telescope. The slit, of length
5\arcmin\  and width 1.3\arcsec, was aligned along the highest surface
brightness
part of the reflection nebulosity, at a position angle of 90\deg~for IC 430,
135\deg~for Re 50 and 45\deg~for L1551 IRS 5. The grating had 316 lines/mm and
was
centered
at 7500\AA; the detector was a Tektronics 2048x2048 CCD. This combination
allowed useful observations to be made from 5500\AA\ to 9500\AA\ with a
spectral resolution of 7\AA. Three
1800 sec observations were made of Re 50 and of L1551 IRS5;
two 1800 sec observations were made
of IC 430. On the night of 1993 January 27 UT we obtained a spectrum of
FU Orionis with the same instrumental setup.

The spectra were reduced using the TWODSPEC package within
IRAF\footnotemark[2]. The two dimensional spectra in the observational
\footnotetext[2] {IRAF is distributed by the National Optical Astronomy
Observatories, which is operated by the Association of Universities for
Research in Astronomy, Inc. (AURA) under cooperative agreement with the
National Science Foundation.}
coordinate system were transformed to a two dimensional array in linear
wavelength and spatial coordinates by making use of the multiplicity of
night sky emission lines crossing the spectra and traces of stellar spectra
placed
at different locations on the slit. This procedure allowed improved sky
subtraction to be achieved in regions where the nebular surface
brightness was low and long slit lengths were coadded.

\section{Discussion}

In Figure 1 we show the H$\alpha$ profiles for Re 50, IC 430, L1551 IRS5,
and FU~Ori. Re~50 and IC~430 show P Cygni H$\alpha$
profiles with the blue wings evidencing extreme outflow velocities,
extending to 1000 km/s for Re 50. The profile morphology is very similar
to that of L1551 IRS 5 which was previously classified as an
FU Ori object by Mundt et al. (1985) on the basis of the H$\alpha$
profile and by Carr, Harvey and Lester (1987) on the basis of the infrared
CO bands. The spectra in both the red region
(5500-7500\AA) and the near-IR region (7700-9200\AA) compare well with that
of FU Ori and a G2 Ia supergiant spectrum taken previously with the KPNO
echelle
and smoothed to our present resolution. The spectra do not
compare well with late G main sequence spectra.  This is shown in Figure
2 where the normalized spectra in the 5600-6800\AA\ region are shown.
Also shown is the region of the infrared Ca triplet, which behaves differently
for each object. While the Ca lines appear in net absorption in the
spectrum of FU Ori, Welty et al. (1992) have shown, by subtracting disk
photosphere model spectra from the observations of FU Ori and V1057 Cyg
that these stars show residual emission in the Ca triplet as well.
In Z CMa, another probable FU Ori object (Hartmann et al. 1989), the Ca triplet
is strongly in emission.
The spectrum of IC 430 shows an emission component to the 8662\AA\ triplet line
while the 8498\AA\ and 8542\AA\
lines show no evidence for emission. The spectrum of IRS 5 shows a similar
phenomenon where the 8662\AA\ line is seen in emission and the two
shorter wavelength lines are not seen, although the higher noise level
from the OH emission in the early evening night sky makes the result of
low significance for the 8498\AA\ line. A discussion of the Ca infrared
triplet anomaly in pre-main sequence stars can be found in Hamann \&
Persson (1992). It is difficult to place these objects within that context
since the emission components are relatively weak; models would have to be
matched to these
spectra and subtracted before a true measure of the emission in each line
could be given, something best done with higher resolution spectra.
Based primarily on the H$\alpha$ profile and apparent low surface gravity and
secondarily on the Ca infrared triplet, we conclude that Re 50
and IC 430 show many of the spectral characteristics similar to those of FU Ori
objects.

The characteristics that are used to define an FU Ori object have evolved
with time. The first identifications were centered on the observation of
a sudden large rise in luminosity over a short period of time. This
allowed 3 objects, FU Ori, V1057 Cyg, and  V1735 Cyg (Elias 1-12)
to be placed in this group.  V1515 Cyg is also included although it has
exhibited a slow rise lasting many decades. It has been recognized however
that this
phenomenon is characteristic of very early stages in a star's evolution,
and is probably associated with multiply occurring sudden mass accretion events
(Herbig
1977). Recent theoretical work (Hartmann, Kenyon \& Hartigan 1993 and
references therein) has modeled the FU Ori phenomenon as a sudden
increase in the mass accretion rate through a circumstellar accretion
disk and has made predictions of spectral features, particularly the 2.3$\mu$m
CO absorption bands, and line profile
doubling to be expected for the disk model.

Because it is not possible to obtain optical spectra of the resolution
necessary to search for line doubling, and we have not yet obtained near
infrared spectra, are there other consistency tests that
can be applied? 1)In images of FU Ori objects, small arctuate nebulosities are
apparent (Goodrich
1987). The reflection nebulosity for both Re 50 and IC 430 is
typical of the form found for FU Ori objects.
2) While the photometric measurements of
V883 Ori are few (2), they are separated by 11 years (Allen et al. 1975,
Nakajima et al. 1986), and offer the possibility of detecting a long
term decline in the brightness of V883 Ori. In this time period, the  object
declined by 0.58
mag at J, 0.57 mag at H, 0.14 mag at K and 0.07 mag at L. Further photometric
monitoring will allow us to confirm the decline of this object.
3) The bolometric luminosities of the known FU Ori objects
range from $\sim$40 $L_{\sun}$(L1551 IRS5) to $\sim$3000 $L_{\sun}$ (Z CMa),
(Rodriguez et al. 1990; Hartmann et al. 1989),
although $\sim\onehalf$ of the luminosity of Z CMa may be attributable to its
infrared companion.
The bolometric luminosity of Re 50 is $\sim$ 300 $L_{\sun}$ while that for IC
430 is $\sim$ 400 $L_{\sun}$. While these are the two most luminous
IRAS sources in L1641, they were separated from the
rest of the high luminosity embedded sources by Strom et al.
(1989) because they did not possess steep red spectral energy
distributions. Instead their spectral energy distributions were
relatively flat. These spectral energy distributions are more
characteristic of the FU Ori objects.
4)~The previously classified FU Ori objects (11 objects,
Kenyon et al. 1993) show optically thin mm continuum emission which
measures the mass of cold dust in the
immediate environment of the object (Weintraub, Sandell, \& Duncan 1989, 1991;
Reipurth et al. 1993) and predominantly located in a
structure with small covering angle, therefore allowing the objects to be
seen.  Both Re 50 and IC 430 have been measured at 1300$\mu$m by Reipurth et
al. (1993). The deduced masses for the circumstellar gas and dust
immediately surrounding these objects are 3.7 $M_{\sun}$ for IC 430 and 1.8
$M_{\sun}$ for Re 50, well above the masses deduced for most of the optically
visible
FU Ori objects but on the low end of the mass distribution for those
objects known to be driving outflows delineated by HH objects.
5) Several FU Ori objects show cm radio emission with a spectral index
characteristic of ionized outflows (Rodriguez, Hartmann \& Chavira 1990,
Rodriguez \& Hartmann 1992)
Both IC~430 and Re~50 were observed in the VLA snapshot survey of Morgan,
Snell and Strom (1990) at 6~cm. Re 50 was clearly detected and the spectral
index demonstrated the emission to be thermal. IC 430 was not detected,
but the 3$\sigma$ noise level is consistent with the range of flux
levels measured for FU Ori objects.
6) Several of these objects are also apparently exciting sources of
HH objects (L1551 IRS5, HH28,29 (Stocke et al. 1988); V346 Nor, HH57 (Reipurth
1985b)), and long narrow jets (Z CMa, Poetzel et al. 1989) indicative of
the energtic outflows diagnosed by the H$\alpha$ profiles.
Both of these objects are also associated with HH objects, HH 65 (Reipurth
1989) with Re 50 and a small emission knot with IC 430 (Strom et al. 1986).
7) Approximately half of the FU Ori objects are known to drive molecular
outflows (see the summary table
in Hartmann, Kenyon and Hartigan 1993). The outflow associated with Re
50 has been well documented (Bally \& Reipurth 1986, Fukui et al. 1986) and its
interaction
with the surrounding cloud material (Casali 1991; Scarrott \&
Wolstencroft 1988) has been described. However, a molecular outflow has
not been found to be associated with IC~430 although a search for such
evidence has been conducted (Morgan, Schloerb, Snell \& Bally 1991).
Neither is there a CS core associated with this object (Tatematsu et al.
1993) although other outflow sources in the cloud are embedded in such
cores.
However, there is other evidence that an energetic outflow however
variable the rate, may have been present in the past. The $^{13}$CO map of the
L1641 molecular cloud (Bally et al.
1987) shows an elliptical depression in the molecular gas distribution,
centered on
this object. The semi-major axis of this cavity is $\sim$ 0.9 parsecs.
The presence of this cavity is also clearly indicated in our 100$\mu$m
IRAS HIRES image of the region. The piled up material at the edges of the
cavity emits strongly at 100$\mu$m.

The presence of this object in the Index Catalog is curious in itself
since the surface brightness of the nebulosity is so low. Therefore we
traced the origin of its listing to a paper by Pickering (1890).
This object is described as a nebulous band $3\arcmin$
wide extending $10\arcmin$ north preceeding from DM-7$\deg$~1142. It is
possible now to see nebulosity of this description on R and I band CCD
images, although we know that the bright star is not responsible for
its illumination. However the surface brightness is only $\sim$20
mag/\sq\arcsec on our R band CCD image. The
photographic plates of that period were sensitive only in the blue. On our
B band CCD image taken 1990 Nov 9 UT, it is very difficult to see the
nebulosity. The maximum surface brightness near V883 Ori is 24 mag/\sq\arcsec
on this image.
It is unlikely that nebulosity of this surface brightness was apparent on
the plates taken in 1888.
Therefore it is probable that V883 Ori was considerably brighter at that time,
and that the epoch of most recent outburst for this object can be placed
in or before 1888, when the two discovery plates were taken. The total
length of this nebulosity, as measured from V883 Ori, is $\sim$1 pc, in
close agreement with the size of the cleared region of the cloud as
seen in $^{13}$CO and 100$\mu$m images.

\section{Conclusions}

Spectra of two heavily embedded IRAS sources in L1641 reveal them to have
characteristics of
FU Ori objects. The other properties of these objects (morphology, bolometric
luminosities and spectral energy distributions, HH objects, inferred disk
masses, thermal radio emission,
and molecular outflow) strengthen
the conclusion that they belong to this class. The infrared CO bands should
be observed to confirm these
identifications. The presence of two FU Ori objects within \onehalf\
 degree of each other in this cloud emphasizes the active star formation
occurring in this cloud at the present time.

\acknowledgements

The authors would like to thank K. Michael Merrill for obtaining the entry
and references in the Index Catalog for us. We would also like to thank
Suzan Edwards for discussions and for leading a successful late night search of
the new Smith College
Science Library for the as yet uncataloged Annals of the Harvard College
Observatory.

\pagebreak

\begin{figure}
\plotfiddle{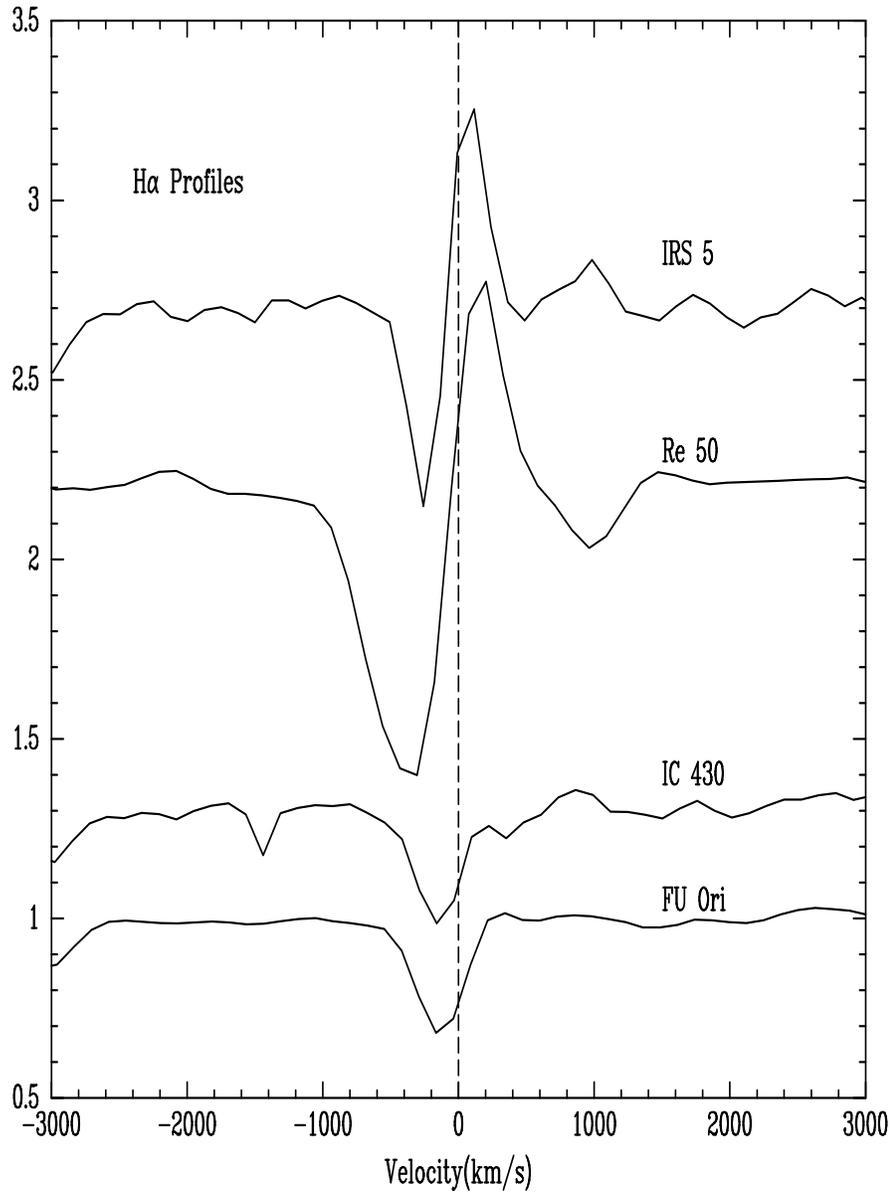}{17.0cm}{0}{65}{65}{-225}{0}
\caption{The H$\alpha$ profiles for FU Ori, IC 430, Re 50 and IRS 5 are shown
on a velocity scale. The P Cygni profiles show little or no emission above
the continuum and blue wings to the absorption components extending to large
outflow velocities.}

\end{figure}
\pagebreak
\begin{figure}
\plotfiddle{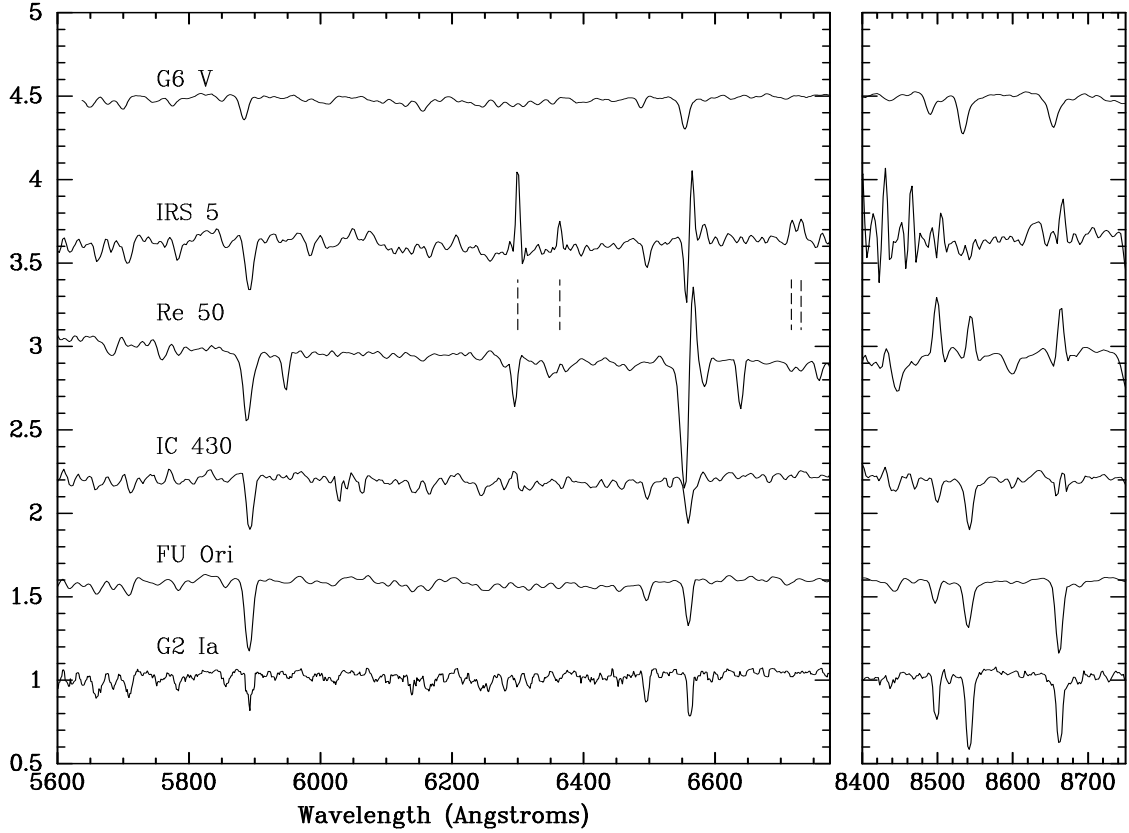}{12.0cm}{-90}{60}{60}{-250}{360}
\caption{Spectra for the FU Orionis objects as well as for a supergiant and
main sequence comparison star. The regions where background emission lines
([O I] and [S II]) have either been incompletely or over subtracted are
marked for IRS 5 and Re 50. The spectra resemble the supergiant standard
and FU Ori more closely than the main sequence star. The Ca triplet
behavior is quite varied, as discussed in the text.}

\end{figure}

\end{document}